# Interface induced room-temperature ferromagnetism in hydrogenated epitaxial graphene

A.J.M. Giesbers[1,*], K. Uhlířová[2], M. Konečný[1,4], E. C. Peters[3], M. Burghard[3], J. Aarts[2], and C.F.J. Flipse[1,†]

[1]*Molecular Materials and Nanosystems, Eindhoven University of Technology, 5600 MB Eindhoven, The Netherlands.*
[2]*Magnetic and Superconducting Materials, Leiden Institute of Physics, 2333 CA Leiden, The Netherlands.*
[3]*Max-Planck Institute for Solid State Research, Heisenbergstrasse 1, D-70569 Stuttgart Germany.*
[4]*CEITEC BUT, Technická 10, 616 69 Brno, Czech Republic*

**Contact information**
[*]A.J.M.Giesbers@tue.nl, [†]C.F.J.Flipse@tue.nl

**Due to the predominantly surface character of graphene, it is highly suitable for functionalization with external atoms and/or molecules leading to a plethora of new and interesting phenomena. Here we show ferromagnetic properties of hydrogen-functionalized epitaxial graphene on SiC. Ferromagnetism in such a material is not directly evident as it is inherently composed of only non-magnetic constituents. Our results nevertheless show strong ferromagnetism, which cannot be explained by simple magnetic impurities. The ferromagnetism is unique to hydrogenated epitaxial graphene on SiC, where interactions with the interfacial buffer layer play a crucial role. We argue that the origin of the observed ferromagnetism is governed by electron correlation effects of the narrow Si-dangling-bond (Si-DB) states in the buffer layer exchange-coupled to localized states in the hydrogenated graphene layer. This forms a quasi-three-dimensional ferromagnet with a Curie temperature higher than 300 K.**

Owing to its capability of ballistic transport over micrometer distances[1], as well as its very long spin relaxation time and spin relaxation length[2, 3], graphene represents a close-to-ideal material for spintronic applications[4]. In this context, considerable effort has recently been directed to rendering graphene ferromagnetic via chemical modification. Thus far, ferromagnetic order in graphene has been attained through covalent functionalization, involving the linkage of radical species like the spin-bearing carbon atom of an organic molecule or hydrogen atoms to the graphene layer[5-17]. Along these lines, functionalization of epitaxial graphene by aryl radicals has been reported to yield disordered magnetism, comprising a mixture of ferromagnetic, superparamagnetic and antiferromagnetic regions[18]. With the aid of combined atomic and magnetic force microscopy, it could be proven that these randomly dispersed regions are constituted by the attached moieties. This lack of a periodic functionalization pattern of the graphene sheet prevents the achievement of long range ferromagnetic order, thus limiting the use of such samples in spintronic devices. Furthermore, room temperature ferromagnetism has been detected in partially hydrogenated epitaxial graphene grown on silicon carbide (SiC), and attributed to hydrogen monomers bonded to the graphene sheet[12]. Despite these accomplishments, however, both the mechanism underlying the ferromagnetic ordering, and the role played by the SiC substrate used for the epitaxial graphene growth, has not yet been clarified. Here, we experimentally demonstrate that spin ordering within hydrogenated epitaxial graphene critically depends on the presence of the underlying buffer layer. In addition, it is shown that the created magnetic areas are distributed over the entire graphene sheet, thus enabling to effectively tune the overall magnetization through the density of attached hydrogen atoms.

To explore the ferromagnetism in epitaxial graphene, we use samples grown on insulating 6H-SiC substrates following the procedure described in ref. 19 (all samples originate from the same wafer). The atomic force micrograph (AFM) of the sample surface (Fig. 1a) reveals terrace steps originating from a slight miscut of the SiC substrate. The terraces are typically 3-5 um wide and approximately 10 nm high and are overgrown with a continuous carpet of graphene[20, 21]. The inset in Fig. 1a shows a schematic cross-section of the layer sequence at the surface with a graphene layer on top of an interfacial carbon layer (buffer layer) partly bonded to the SiC substrate. On the terrace edges an unintentional region of bilayer graphene has formed under the current gowth conditions[19], discernible as brighter areas in the corresponding AFM phase image (Fig. 1b). The presence of a small bilayer area is confirmed by Raman microscopy and low energy electron microscopy (LEEM) investigations (see supplementary information Fig. S1) and has the same coverage in all samples. After growth, the graphene samples are hydrogenated by an atomic hydrogen source in an ultrahigh vacuum chamber, for different exposure times. Successful hydrogenation is testified by an enhancement of the $sp^3$-defect associated Raman D-peak, whose intensity increases with treatment time (Fig. 1c and d), as discussed in more detail in the supplementary information. Increasing hydrogen exposure also leads to a rising C-H signal in x-ray photo-absorption spectra (XPS) (see supplementary information). The inset in Fig. 1d illustrates the hydrogen bonded on the top graphene layer.

The magnetic properties of the hydrogenated graphene samples are determined using a commercial SQUID with a sensitivity of $5 \cdot 10^{-8}$ emu. All measurements are performed at room temperature unless stated otherwise. Figure 2a shows the magnetization of an epitaxial graphene sample hydrogenated for 3 minutes. The linear background is related to the bulk SiC diamagnetism and can be subtracted by a linear fit to the high field part of the curve where all other forms of magnetism are assumed to be saturated. The resulting diamagnetic susceptibility $\chi = \mu_0 M/Hm$, with $m = (1.97 \pm 0.05) \cdot 10^{-5}$ kg the sample mass and $\mu_0 = 4\pi \cdot 10^{-7}$ Tm/A the vacuum permeability, is $\chi_{SiC} = -(4.1 \pm 0.1) \cdot 10^{-9}$ m$^3$/kg, within its error in good agreement with literature ($\chi_{SiC} = -4.01 \cdot 10^{-9}$ m$^3$/kg). Consistent values for $\chi_{SiC}$, within the error range, were found for all samples used in this work. The data obtained after subtraction of the SiC diamagnetic background are shown in Fig. 2b for three different temperatures. The curves show a clear ferromagnetic response from the hydrogenated epitaxial graphene. The hysteresis loop shows a saturation magnetization of $M_s = \pm 27 \cdot 10^{-7}$ emu, a remanent magnetization of $M_r = \pm 7 \cdot 10^{-7}$ emu and a coercive field of $H_c = \pm 91$ Oe at 300 K. Upon decreasing the temperature a small increase in the high field magnetization occurs. A similar trend is observed for the coercive field and the remanent magnetization (inset Fig. 2b). The measured saturation magnetization at room temperature corresponds to a value of about $0.9 \mu_B$ per unit cell.

Figure 2c compares the ferromagnetic signal for the 3 min hydrogenated sample under in-plane magnetic field, along ($\theta = 0$ deg) and perpendicular ($\theta = 90$ deg) to the terraces, as well as for out-of-plane (OofP) orientation (inset Fig. 2c). A notable anisotropy can be discerned, with easier magnetization along the terrace steps (black curve), as compared to perpendicular alignment (red curve) and the out-of-plane direction (blue curve). This difference manifests itself in a lower saturation magnetization and in the case of the out-of-plane signal in a more stretched hysteresis loop. The preferred magnetization along the terrace edges might result from the predominant formation of double site hydrogen sites aligned along the zigzag direction of graphene[22]. The double H-sites show elongated shaped charge structures of 3 nm or more with 6-fold symmetry coinciding with the 6 fold symmetry of the graphene honeycomb lattice. In atomic resolution STM it was shown that the armchair edge of the graphene layer coincides with the SiC terrace structure[23]. Combined, these results lead to anisotropy between the terrace edge and perpendicular to the edge direction which could explain the observed anisotropy in the magnetization. The out of plane magnetization contribution is probably due to a non collinear spin orientation in the buffer layer, similar as for the $\sqrt{3} \times \sqrt{3} R30$ 6H-SiC(0001) structure of SiC[24], which will be discussed later.

To tune the ferromagnetic signal we can use the hydrogen coverage, as is shown in Fig. 2d for hydrogenation times between 0 and 120 minutes. While the pristine graphene (0 min, black curve) displays no magnetic signal, a short hydrogen exposure (0.5 min, red curve) results in a clear ferromagnetic signal. From the corresponding hysteresis loop, a coercive field of $H_c = \pm 65$ Oe and remanent magnetization of $M_r = \pm 2.4 \cdot 10^{-7}$ emu is extracted. At high fields ($H = 3000$ Oe), the magnetization reaches a saturation value of $14 \cdot 10^{-7}$ emu. This saturation magnetization, $M_s$, increases up to a treatment time of 3 minutes ($27 \cdot 10^{-7}$ emu), which is followed by a decrease for longer treatments, finally resulting in $M_s$ (120 min) = $13 \cdot 10^{-7}$ emu. The same trend is observed in the coercive field and the remanent magnetization for the different samples.

In order to determine the origin of the ferromagnetic behavior, we have investigated the magnetization properties of several control samples (Fig. 2c). Firstly, a sample prepared in the same manner as the 3 minute sample, except that the hydrogen bottle is kept closed, is found to exhibit no ferromagnetic signal (red curve). Secondly, the same procedure is applied to an untreated bare SiC sample, which likewise does not lead to ferromagnetic signatures (not shown). Thirdly, to test the influence of the underlying substrate, a quasi-freestanding monolayer of graphene[24] (QFMG) is used as a third control sample (Fig. 3f shows a schematic). It is obtained by growing only a buffer layer[24-26] on the SiC, followed by hydrogen intercalation to passivate the SiC substrate and turn the buffer layer into a QFMG. Owing to the reduced substrate interaction, QFMG is of superior quality compared to epitaxial graphene[24]. Pristine QFMG samples not subjected to hydrogenation (exemplified by green curve) do not display ferromagnetism as expected for pure graphene (in total two samples were studied). Remarkably, also after 3 minutes of hydrogenation, no ferromagnetic signal at room temperature emerges for such samples (hQFMG, schematic in Fig. 3f) (blue curve representative for one out of two samples). The above findings highlight that the hydrogenated graphene is not ferromagnetic at room temperature and the buffer layer is crucial to render the epitaxial graphene ferromagnetic. Finally, in a fourth control experiment using two buffer layer samples (one shown, cyan curve) and three hydrogenated buffer layer samples (one shown, magenta curve) no ferromagnetic signal is detected (schematics of the samples are shown in Fig. 3f). This absence consolidates that hydrogenated epitaxial graphene requires both the hydrogenated graphene and the underlying buffer layer to become ferromagnetic. At low temperatures, the linear background magnetization, observable for both the buffer layer samples and the hQFMG, leads to a smaller $\chi_{SiC}$ compared to the pure SiC substrates. This difference hints toward an unsaturated low temperature paramagnetic contribution in these samples, akin to fluorinated graphene laminates[27]. The presence of localized paramagnetic-like states in the buffer layer was recently also suggested from spin transport experiments in epitaxial graphene[28]. In our preliminary high magnetic field magnetization measurements the paramagnetism of the bufferlayer is indeed confirmed, saturating at $H/T \approx 25$ kOe/K (see supplementary information Fig S4).

Further insight into the ferromagnetic properties of the hydrogenated graphene is gained by detecting the remanent magnetization with the aid of Magnetic Force Microscopy (MFM) (see Fig. 3). By placing the sample briefly on either the south pole (-B) or north pole (+B) of a permanent magnet prior to MFM measurements, we can magnetize the sample in respectively a negative or positive out-of-plane remanent magnetization state as we observed in the SQUID measurements of Fig. 2c. Figure 3a and b show the magnetic signal of the same area for the two magnetization directions with their respective cross-sections in panel c. The highlighted dirt particle is an artifact due to crosstalk with the topography[29] and serves as a position marker on the sample. The topology of the sample is similar to Figure 1a and b. The clear difference in MFM contrast between the single (1L) and bilayer (2L) areas, indicate their different magnetization. This might be due to different hydrogen coverages[30, 31], in accord with the lower overall D-peak intensity on the bilayer regions observed in Raman images. Other possible contributions are the different electronic structure of the bilayer graphene, as well as different interactions among the hydrogen sites, or in the specific case of the bilayer graphene between hydrogen sites and the buffer layer due to the increased distance between the buffer layer and the hydrogenated layer. In the SQUID measurements the bilayer areas will reduce the overall saturation magnetization, however since the bilayer coverage is similar for all samples the results above are not affected. The switching of the out-of-plane remanent magnetization direction is clearly visible in the MFM cross-sections in Fig. 3c. Specifically, after positive B-field magnetization, the MFM signal is positive and the signal from the single layer is slightly larger than that from the bilayer. After negative B-field magnetization the MFM signal has reversed the sign and the response from the single layer is again highest. These changes show that the color inversion between panel a and b is due to a complete flip of the magnetization direction, while the signal from the single layer is always higher than that from the bilayer. That the flip is not symmetric around zero indicates a constant background phase shift, and might be attributed to electrostatic interactions simultaneously probed by the metallic tip. Electric field microscopy (EFM) confirmed this magnetic-field independent electrostatic background[32] (see supplementary figure S3).

The MFM measurements corroborate the ferromagnetism of the hydrogenated epitaxial graphene sample and show that the signal originates from the whole surface. Together with the observed variation of the ferromagnetic strength with hydrogen coverage, the magnetic anisotropy and control sample magnetic measurements these results form a conclusive set of observations which rule out any possible magnetic contaminations as the origin of the observed magnetic behavior.

The observed ferromagnetism in our hydrogenated epitaxial graphene is best interpreted in terms of an exchange coupled interaction between localized electron states of the buffer-layer and either spin-polarized localized states or the mid-gap states of the hydrogenated graphene layer[22]. The overall paramagnetic behavior of the buffer-layer indicates the presence of localized magnetic moments, which are the localized defect states forming an insulating behavior as has been shown by STM and STS experiments[33, 34]. These states are attributed to Si-dangling bond (DB) states which probably behave as Hubbard Coulomb repulsion driven Neel like states, described by a non-collinear spin density wave, similar as was shown by Anisimov[35] for a smaller unit cell reconstructed surface, the √3x√3R30 6H-SiC(0001). The SiC buffer-layer has a 6 times larger unit cell, 6√3x6√3R30 SiC(0001) surface structure with a band gap of 1 eV, formed by localized Si-DB states[33, 34]. Upon hydrogen adsorption on top of the graphene layer, carbon-hydrogen bonds are created, forming a mid-gap state[22]. This localized mid-gap state can be spin-split in filled and unfilled localized states close to the Fermi-level due to the Coulomb interaction of the Si-DB states of the buffer layer, forming a quasi 3-dimensional ferromagnetic state with a Curie temperature of 300 K or higher. A second option to explain the FM behavior at 300 K is that the hydrogenated graphene layer is intrinsically ferromagnetic, but with a much lower Curie temperature due to the two-dimensionality, which would become quasi-three dimensional if the paramagnetic buffer-layer will exchange couple to it.

To conclude, hydrogenated epitaxial graphene shows a ferromagnetic behavior with a Curie temperature higher than 300 K and a magnetic moment of $0.9\mu_B$ per effective carbon hexagon area. We have shown that both the hydrogen coverage and the buffer-layer with the Si-dangling bonds, play a crucial role for the high temperature ferromagnetic properties. To explain the ferromagnetism in our graphene system at room temperature, we tentatively propose an exchange coupled interaction between the Coulomb induced localized Si-DB states of the buffer-layer and the localized mid-gap state or the two-dimensional ferromagnetic hydrogenated graphene layer. The buffer-layer stabilizes the ferromagnetic behavior at room temperature and this quasi three-dimensional system can explain the relatively high Curie temperature, higher than 300 K. The high Curie temperature in combination with a small coercive field (100 Oe) and high spin relaxation time in graphene makes hydrogenated epitaxial graphene a favorable material for spintronic applications.

**Methods**

For the magnetization measurements we used a Magnetic Properties Measurement System from Quantum Design with Reciprocating sample magnetometer. The best resolution was obtained with an amplitude of 3 cm and a frequency of 2Hz with a system resolution of $5 \cdot 10^{-8}$ emu. The samples were placed into standard ø 6 mm PE straws which exactly fixated the 4x4 mm sample without further adhesives.

The magnetic signal in the MFM measurements is acquired by a magnetic Co-Cr coated tip with a force constant of $k = 2.8$ N/m and a quality factor $Q = 226$ in constant lift mode of 50 nm and an oscillation amplitude of 30 nm is used to avoid short range and van der Waals interactions. The MFM signal, or phase-shift $\Delta\theta$, is directly related to the magnetic force $F$ between the tip and substrate:

$$\Delta\theta \approx -\frac{Q}{k}\frac{dF}{dz},$$

leading to a force in the order of nano-Newtons.

**Figures:**

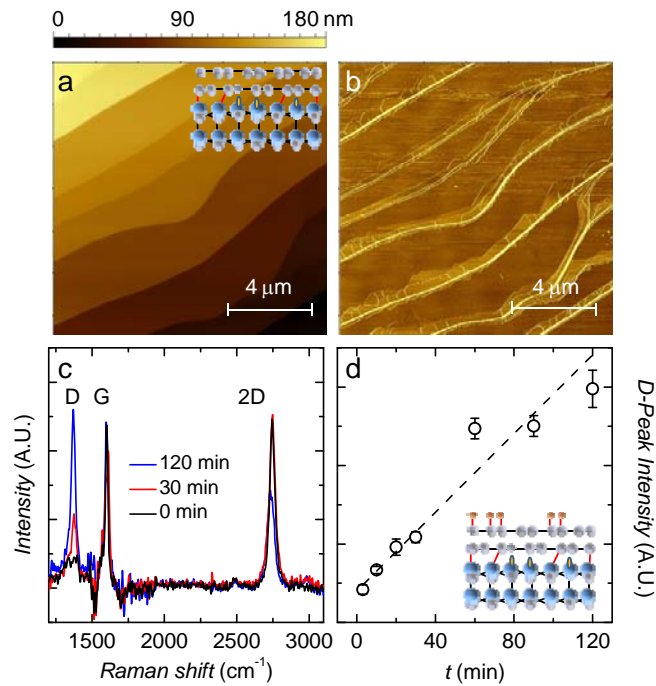

**Figure 1| Epitaxial graphene on 6H-SiC.** (a) Atomic force micrograph of the typical SiC terrace structure on top of which the graphene is grown. The inset shows a schematic of the surface structure with from top to bottom a graphene layer, a buffer layer and the SiC substrate. The gray spheres represent carbon, the blue spheres silicon, and the yellow ovals the silicon dangling bonds. (b) Phase image showing the single layer graphene areas on top of the terraces and the narrow bilayer regions at the terrace edges. (c) Raman spectra of hydrogenated graphene with a treatment time $t = 0$ min. (black), 30 min. (red) and 120 min. (blue). Clearly visible is the upcoming D-peak intensity with increasing treatment time. (d) D-peak intensity as a function of treatment time (the line is a guide to the eye). The inset shows the schematic bonding of hydrogen (orange spheres) to the graphene layer.

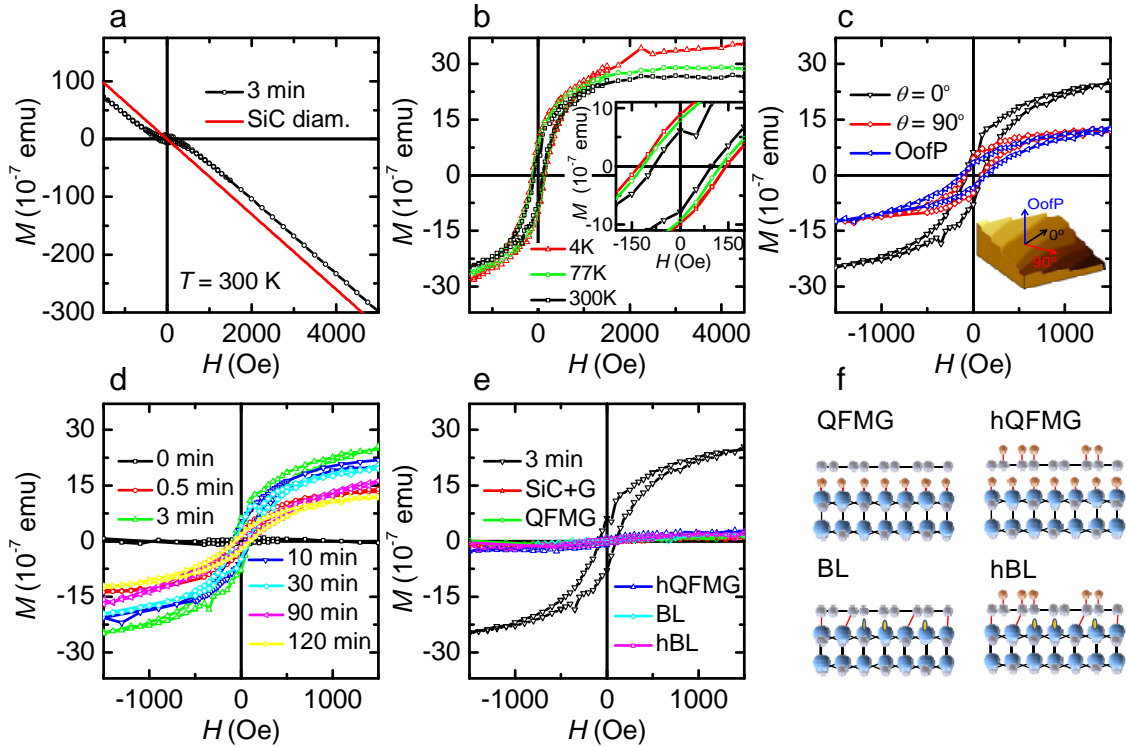

**Figure 2| Magnetization of hydrogenated epitaxial graphene.** (a) Room temperature magnetization as a function of the applied magnetic field for hydrogenated epitaxial graphene treated for 3 minutes. The red line shows the diamagnetic contribution of the SiC substrate. (b) Temperature dependence of the magnetization after subtraction of the diamagnetic background showing a clear ferromagnetic hysteresis loop. The inset shows a zoom of the coercive field and remanent magnetization (3 min. treatment). (c) Direction dependence of the magnetization (3 min. treatment). (d) Ferromagnetic signal for different hydrogen treatment times, $t$ = 0, 0.5, 3, 10, 30, 90, and 120 minutes. (e) Magnetization for different control samples. (f) Schematic representations of the various control samples.

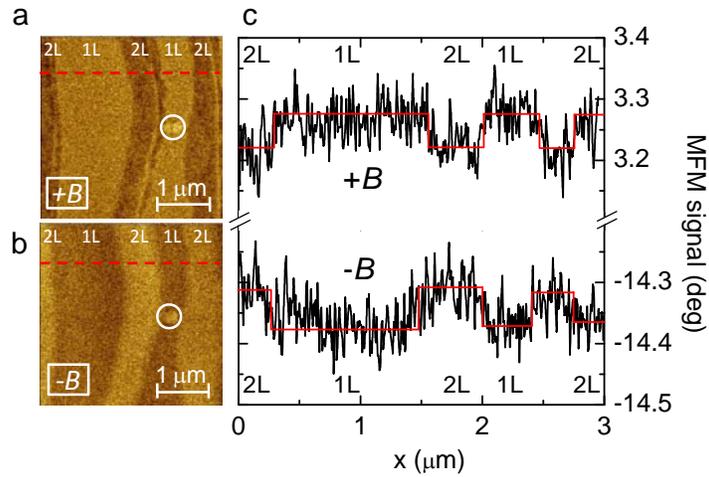

**Figure 3| Magnetic Force Micrographs of hydrogenated graphene.** (a) Magnetic force micrograph after applying a positive magnetic field to the sample showing high and low remanent magnetization for single and bilayer respectively (scale: +3.3 ± 0.2 deg). (b) Inversion of the remanent magnetization after applying a negative magnetic field to the sample (scale: -14.4± 0.2 deg). (c) Cross-section of the positive (a) and negative (b) magnetization. We repeated the switching between positive and negative magnetization several times yielding the same result.

**Supplementary information**

**Samples**

The epitaxial graphene samples used in this research are grown on insulating 6H-SiC substrates from II-VI Inc. following the procedure described in ref. [S1]. All samples in this article originate from the same wafer from II-VI inc. (EI1037-07-EV) that was diced in 4x4 mm pieces by laser cutting. After growth the sample surface consists of 1-5 µm wide terraces with a step height of about 10 nm as was shown by AFM in the main text. The terraces are covered with single layer graphene with a narrow strip of bilayer graphene at the edges. In the main text this was visualized by the AFM phase image, figure S1 shows a typical Raman map of the surface confirming the alternating structure by plotting the 2D-peak area and 2D-peak position, respectively Fig. S1a and b. The dark red areas indicate single layer graphene and the bright yellow areas indicate bilayer graphene as can be seen from the complete spectra at the red and blue dot displayed in Fig. S1c with matching colors. Figure S1d shows a low energy electron microscopy (LEEM) image of a similar sample again showing the alternating single and bilayer area. The respective intensity versus energy spectra taken on the numbered areas are shown in Fig. S1e. The number of dips in the spectra corresponds to the number of layers.

**Hydrogenation of graphene**

After growth the samples are covered with hydrogen by an atomic hydrogen source in an ultra high vacuum chamber. The hydrogenation is done at a pressure of $2 \cdot 10^{-6}$ mbar and different coverages are achieved by varying the exposure time. After hydrogenation the Raman signature of the sample has drastically changed. Figure 1c (main text) shows the Raman spectrum before and after hydrogenation with a treatment time of 30 minutes and 120 minutes. The main difference between the hydrogenated spectra and the non-hydrogenated spectra is the appearance of an additional peak at 1375 cm$^{-1}$ [S2], the so-called D-peak. The D-peak is caused by breathing modes of sp$^2$ rings and is activated by a defect [S3]. The chemisorption of hydrogen on graphene is expected to create sp$^3$ bonds. These sp$^3$ bonds act as defects and allow sp$^2$ breathing modes, which show up as a D-peak in the Raman spectrum. The intensity of the D-peak is directly related to the amount of defects present [S3], in our case the amount of hydrogen. Figure 1d (main text) shows the evolution of the D-peak intensity as a function of treatment time, which clearly shows the increased hydrogen coverage with treatment time. Recently it was phenomenologically shown [S4] that the intensity ratio of the D and D' peak (at 1620 cm$^{-1}$) for sp$^3$ defects was ~10 or larger. In our SiC-graphene samples the SiC background [S1] impedes a proper analysis of the D' peak, however we repeated the same hydrogen treatments on graphene flakes on SiO$_2$ which indeed showed a I(D)/I(D') ratio of 11, corroborating that we indeed have hydrogenated our samples.

**X-ray photo-absorption spectroscopy**

In addition, the presence of hydrogen is confirmed by X-ray photo-absorption spectroscopy (XPS) measurements (see Fig. S2a-c) [S5-S7]. Figure 2a shows the XPS spectrum for an untreated sample (black line). In accordance to ref. [S5] the spectra can be fitted with four main peaks, one originating from the SiC substrate (red), two from the two differently bonded carbon atoms in the bufferlayer (S1: magenta, S2: cyan) and one from the graphene layer (green). After the sample is treated with hydrogen the main left peak starts to shift to higher binding energies. This shift can be accounted for by adding a new peak for the binding energy of carbon-hydrogen bonds at 284.74 ± 0.10 eV (brown peak in Fig. S2b and c) [S6]. The intensity of this C-H peak increases with increasing treatment time, whereas the C-C peak intensity decreases as expected for higher hydrogen coverage.

**Electric Force Microscopy**

To rule out any electrostatic contributions to the MFM signal in the main text these interactions between a metallic AFM tip and the hydrogenated graphene sample, are measured by electric force microscopy (EFM). Figure S3a shows the AFM topography of the sample measured simultaneously with the EFM signal. Panel b and c show the EFM signal after the sample was place on respectively the south pole (-B) and the north pole (+B) of a permanent magnet. Neither of the EFM measurements shows a response to the magnetic field direction. This indeed confirms that the magnetic field dependent MFM results are related to the remanent magnetization of the hydrogenated epitaxial graphene. The electrostatic interactions with the metallic tip only lead to a constant offset in the MFM measurements as was indeed shown in the main text.

**Paramagnetism in the buffer layer**

To confirm if the smaller diamagnetic background (smaller than $\chi_{SiC}$), measured for the buffer layer, is related to a paramagnetic signal in this layer we perform a high magnetic field magnetization measurement at low temperatures (see Fig. S4). After subtraction of the diamagnetic sample substrate background ($\chi_{SiC}$= -(4.1 ± 0.1)·10$^{-9}$ m$^3$/kg) a clear paramagnetic signal becomes visible. As a function of $H/T$ both the 2.5 K and 4 K curve coincide and can be fitted with a single Brillouin function

$$M = NgJ\mu_B \left[ \frac{2J+1}{2J} ctnh\left(\frac{(2J+1)z}{2J}\right) - \frac{1}{2J} ctnh\left(\frac{z}{2J}\right) \right]$$

with $z = gJ\mu_B H/k_B T$, $N$ the number of spins, $g$ the g-factor, $J$ the angular momentum quantum number, $\mu_B$ the Bohr magneton, $H$ the magnetic field, $k_B$ the Boltzman constant and $T$ the temperature. The best fit was obtained for $J = 1/2$ with $N$ and $g$ as fitting parameters leading to $N = (3.7 \pm 0.3) \cdot 10^{14}$ spins and $g = 2.7 \pm 0.2$. An enhancement of the g-factor of the buffer layer was already shown by Maassen et al. [S8] but has also been observed in graphene [S9]. The extracted number of spins $N$ on our 4 mm × 4 mm sample is equivalent to 1.2$\mu_B$ per hexagon area and can originate from the Silicon dangling bonds [S10] but also from carbon vacancies [S11] and other defect structures present in the unscreened buffer layer.

**References supplementary information**

**Supplementary figures**

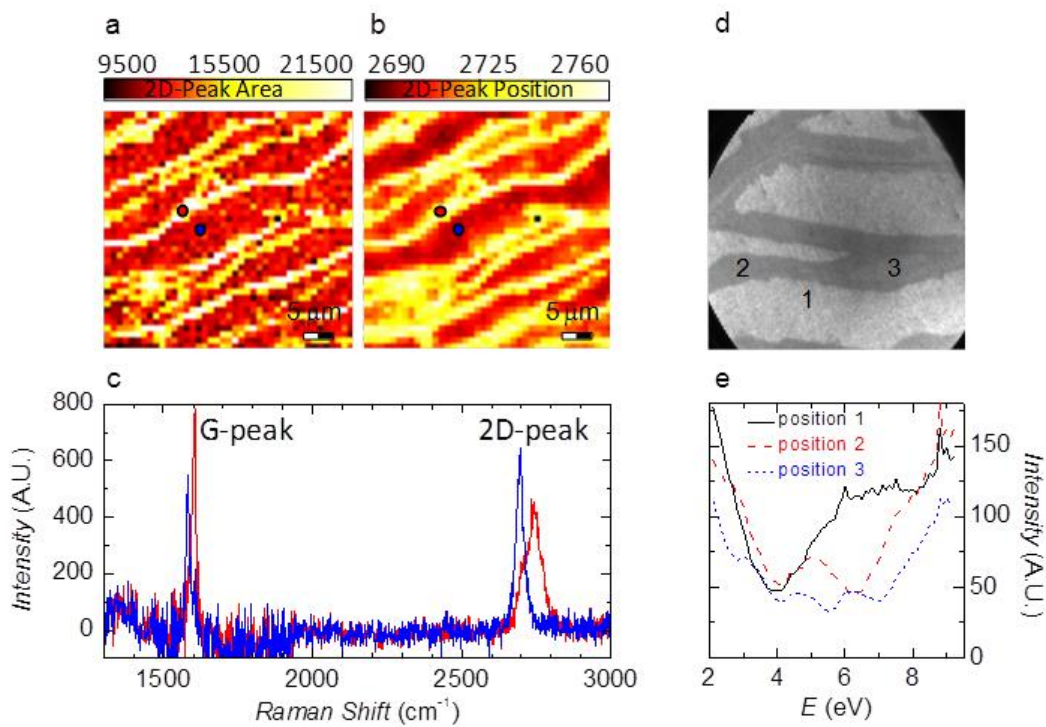

**Figure S1| Number of graphene layers.** Raman map image of the 2D-peak Area (a) and Position (b) showing the alternating single (red) and bilayer (yellow) areas. (c) Full Raman spectra at the red and blue locations indicated in (a) and (b). In blue the typical spectra for a single layer graphene and in red for the bilayer graphene. (d) LEEM image showing a clear contrast difference between the single layer graphene in the middle of a terraces and bilayer graphene on the terrace edges. (d) LEEM intensity as a function of energy on the positions as indicated in (d). The number of dips corresponds to the number of layers, corroborating the alternating single and bilayer areas with incidental multilayer patches.

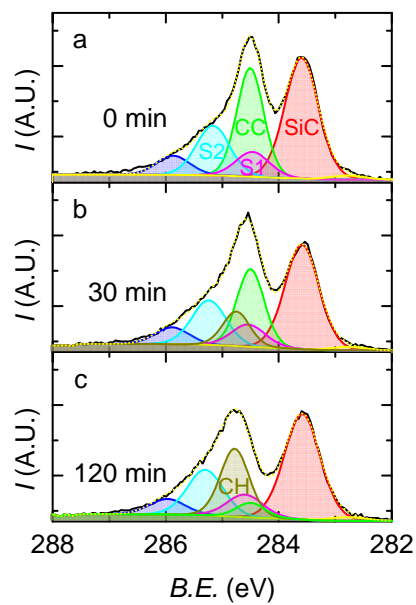

**Figure S2| X-ray photoemission spectra for hydrogenated graphene.** (a) spectra for an untreated epitaxial graphene samples with the S1 (magenta) and S2 (cyan) peaks originating from the bufferlayer, the SiC (red) peak from the substrate and the CC (green) peak from the $sp^2$ graphene carbon atoms. (b) spectrum with a hydrogen treatment of 30 min showing the upcomming CH (brown) peak of $sp^3$ carbon-hydrogen bonds. (c) Same spectra after 120 min hydrogen treatment.

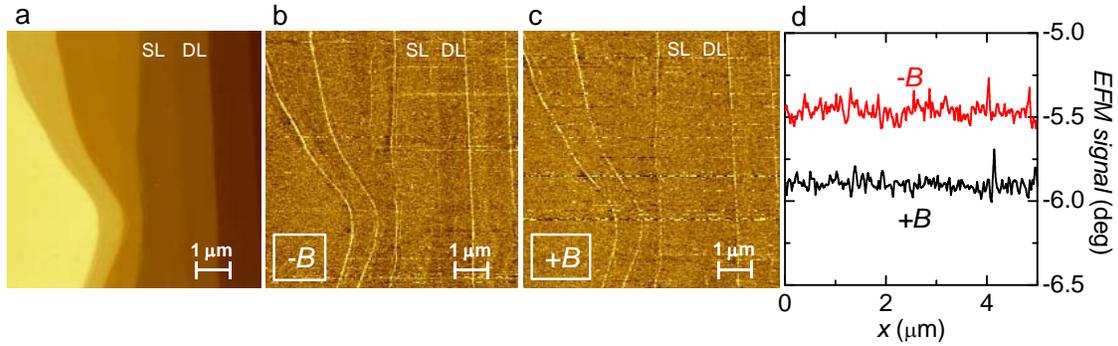

**Figure S3| Electric Force Microscopy of hydrogenated graphene.** (a) Atomic force microscope height image taken with a Pt/Ir conductive tip. (b) Electrical force image after the sample was placed on the south pole of a permanent magnet. (c) Electrical force image after the sample was placed on the north pole of a permanent magnet. (d) Cross section of the images in (b) and (c) after respective negative (-B) and positive (+B) magnetization showing no magnetic field response.

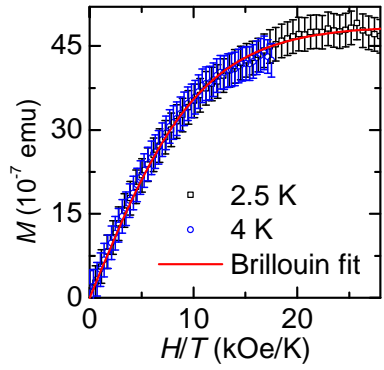

**Figure S4| Paramagnetism in bufferlayer on SiC.** Magnetization of the buffer layer for 2.5 K and 4 K. The curves overlap as a function of H/T after subtraction of the diamagnetic background and show a clear paramagnetic signal saturating at high magnetic fields. The red line is a fit to the Brillouin function for $J = 1/2$.